\documentclass[aps,prb,reprint,showpacs,floatfix]{revtex4-1}  
\usepackage{amsmath,amssymb,xcolor}
\usepackage[colorlinks,bookmarks=false,citecolor=red,linkcolor=blue,urlcolor=blue]{hyperref}
\usepackage{graphicx,epstopdf}
\usepackage{hyperref}
\hypersetup{
    colorlinks,
    citecolor=blue,
    filecolor=blue,
    linkcolor=blue,
    urlcolor=blue}

\begin{document}
\title{ Magnetisation plateaux of the quantum pyrochlore Heisenberg antiferromagnet}
\author{Santanu Pal}
\email{sp13rs010@iiserkol.ac.in}
\author{Siddhartha Lal}
\email{slal@iiserkol.ac.in}
\affiliation{Department of Physical Sciences, Indian Institute of Science Education and Research-Kolkata, W.B. 741246, India}
\begin{abstract}
We predict magnetisation plateaux ground states for $S=1/2$ Heisenberg antiferromagnets on pyrochlore lattices by formulating arguments based on gauge and spin-parity transformations. We derive a twist operator appropriate to the pyrochlore lattice, and show that it is equivalent to a large gauge transformation. Invariance under this large gauge transformation indicates the sensitivity of the ground state to changes in boundary conditions. This leads to the formulation of an Oshikawa-Yamanaka-Affleck (OYA)-like criterion at finite external magnetic field, enabling the prediction of plateaux in the magnetisation versus field diagram. We also develop an analysis based on the spin-parity operator, leading to a condition from which identical predictions are obtained of magnetisation plateaux ground states. Both analyses are based on the non-local nature of the transformations, and rely only on the symmetries of the Hamiltonian. This suggests that the plateaux ground states can possess properties arising from non-local entanglement between the spins. We also demonstrate that while a spin-lattice coupling stabilises plateaux in a system of quantum spins with antiferromagnetic exchange, it can compete with weak ferromagnetic spin exchange in leading to frustration-induced magnetisation plateaux.
\end{abstract}
\pacs{}
\maketitle
\section{Introduction}
Geometrically frustrated lattices are widely expected to harbour exotic states of matter, including quantum spin-liquids~\cite{lee2008end,fu2015evidence,yan2011spin}, spin-ice~\cite{PhysRevLett.79.2554,bramwell2001spin,balents2010spin,
PhysRevLett.116.177203}, fractional excitations~\cite{PhysRevLett.86.1335,han2012fractionalized,balents2010spin} and  magnetisation plateaux~\cite{PhysRevLett.93.197203,PhysRevLett.94.047202,
nishimoto2013controlling,PhysRevB.88.144416} at finite external magnetic field. A plateau in the magnetisation needs the existence of a finite spectral gap, and can sometimes involve a magnetic ground state with non-trivial entanglement~\cite{PhysRevB.90.174409,10.1088/1367-2630/ab05ff,PhysRevB.93.060407}. In keeping with this, a large number of theoretical and experimental studies have sought such exotic states on highly frustrated lattices like the kagome in two spatial dimensions (2D) and the pyrochlore in three dimensions (3D)~\cite{PhysRevX.4.011025,PhysRevLett.110.207208,PhysRevX.7.031020,PhysRevB.84.020407,PhysRevLett.101.117203,RevModPhys.82.53,PhysRevB.97.144407,PhysRevMaterials.1.071201,PhysRevB.78.180410,PhysRevB.79.144432,ravi2018pyrochlore}. 
While studies of the $S=1/2$ kagome antiferromagnet at finite magnetic field have predicted as well as verified the existence of several magnetisation plateaux~\cite{chen2018thermodynamics,pal2018non,PhysRevLett.114.227202}, the pyrochlore counterpart is much less studied. A notable study for the pyrochlore lattice involves a semi-classical (vector spin) symmetry-based analysis by Penc \emph{et al.}~\cite{PhysRevLett.93.197203}. There, the authors showed that a spin-lattice coupling (SLC) may stabilize the $1/2$- magnetisation plateaux state. Such a plateau has been confirmed by recent experiments on the spinel (CdCr$_2$O$_4$) with $S=3/2$ spins on the pyrochlore lattice formed by the network of the Cr sites~\cite{PhysRevLett.94.047202,Kojima2010}. Further, there are indications of magnetisation plateaux in some recent experiments on spin-ice pyrochlore (A$_2$B$_2$O$_7$) systems with large easy-axis anisotropy as well~\cite{borzi2016intermediate}.  
\par 
Quantum interference effects in condensed matter physics have played a special role in elucidating the properties of topological state of matter~\cite{tanaka2015short}. Following the celebrated work of Haldane for ferromagnetic spin chains~\cite{PhysRevLett.57.1488} as well as Tanaka \emph{et al.} on antiferromagnetic spin chains~\cite{PhysRevB.79.064412}, the Euclidean path integral approach has been employed for the calculation of non-trivial geometric phase factors (if any) in the probability amplitude for excitations above the ground state. This involves applying a gradual twist to the real-space order parameter in a system with periodic boundary conditions. As shown by Haldane, the topological quantisation of such geometric phases can give rise to the gapping of the spectrum for the integer spin Heisenberg chains~\cite{PhysRevLett.57.1488}, while the spectrum of the half-integer spin chains remains gapless. Similar conclusions for half-integer spin chains can also be obtained from a twist operator-based argument~\cite{lieb1961two,PhysRevLett.78.1984} which relies on the sensitivity of the ground state to changes in boundary conditions. Such twist operations are equivalent to large gauge transformations that involve the adiabatic insertion of an Aharanov-Bohm flux through the system~ \cite{PhysRevB.23.5632,PhysRevB.30.1097}.
\par 
In following this line of thought, one can define twist and translation operators for the pyrochlore lattice in order to understand qualitatively the ground state properties of the system. Although pyrochlore is a 3D lattice, following works by Oshikawa~\cite{PhysRevLett.84.1535}, Hastings~\cite{PhysRevB.69.104431} and others \cite{PhysRevB.70.245118,nachtergaele2007multi}, we know that the Lieb-Schultz-Mattis (LSM) argument~\cite{lieb1961two} can be extended to higher dimensional systems with short-ranged interactions. In general, applying the LSM argument in higher dimensions ($D>1$) gives the energy of the variational twisted state as $\mathcal{O}(C/L)$ (where $CL$ is the volume, and $L$ is the length of the direction being twisted). Clearly, this energy is not small in the thermodynamic limit in a spatially isotopic system. By considering a strongly anisotropy limit such that $C/L\rightarrow 0$~\cite{PhysRevB.37.5186}, one can then apply the LSM-argument once more. However, by relating the twist operator with a large gauge transformation, Oshikawa~\cite{PhysRevLett.84.1535} showed that taking the LSM argument is valid well beyond the strong anisotropy regime. Recent works have also extended the validity of the theorem to frustrated quantum spin systems~\cite{Nomura,pal2018non}. This needs, for instance, a careful definition of the twist operator by taking into consideration the symmetries of the geometrically frustrated lattice~\cite{pal2018non}.
\par 
In this work, our main aim is to develop a symmetry-based analysis towards predicting the possible existence of several fractional magnetisation plateaux states in the $S=1/2$ pyrochlore system. Following Ref.(\cite{pal2018non}), we extend the twist-operator formalism in deriving an Oshikawa-Yamanaka-Aflleck (OYA)-like criterion for the pyrochlore lattice. Besides this, we develop a spin-parity operator based analysis~\cite{PhysRevB.57.8494,jalal2016topological} of the system, and obtain predictions of magnetisation plateaux identical to those found from the twist-operator method. This identifies spin-parity as a good quantum number for the identification of plateaux states. The paper is organised as follows. In Section II, we discuss the symmetries of the Hamiltonian and show how a SLC renormalises the Heisenberg exchange constant. In Section III, we develop a twist-operator formalism and thereby derive an OYA-like criterion for possible plateaux states of the pyrochlore lattice. Section IV is devoted to the formulation of a spin-parity based criterion for magnetisation plateaux, and a comparison made with those obtained from the twist-operator method. We conclude in Section V with discussion of the results and some future directions. Details of some of the calculations are provided in the Appendices. 
\section{Hamiltonian for the pyrochlore lattice}
\begin{figure}[h!]
\centering
\includegraphics[scale=.4]{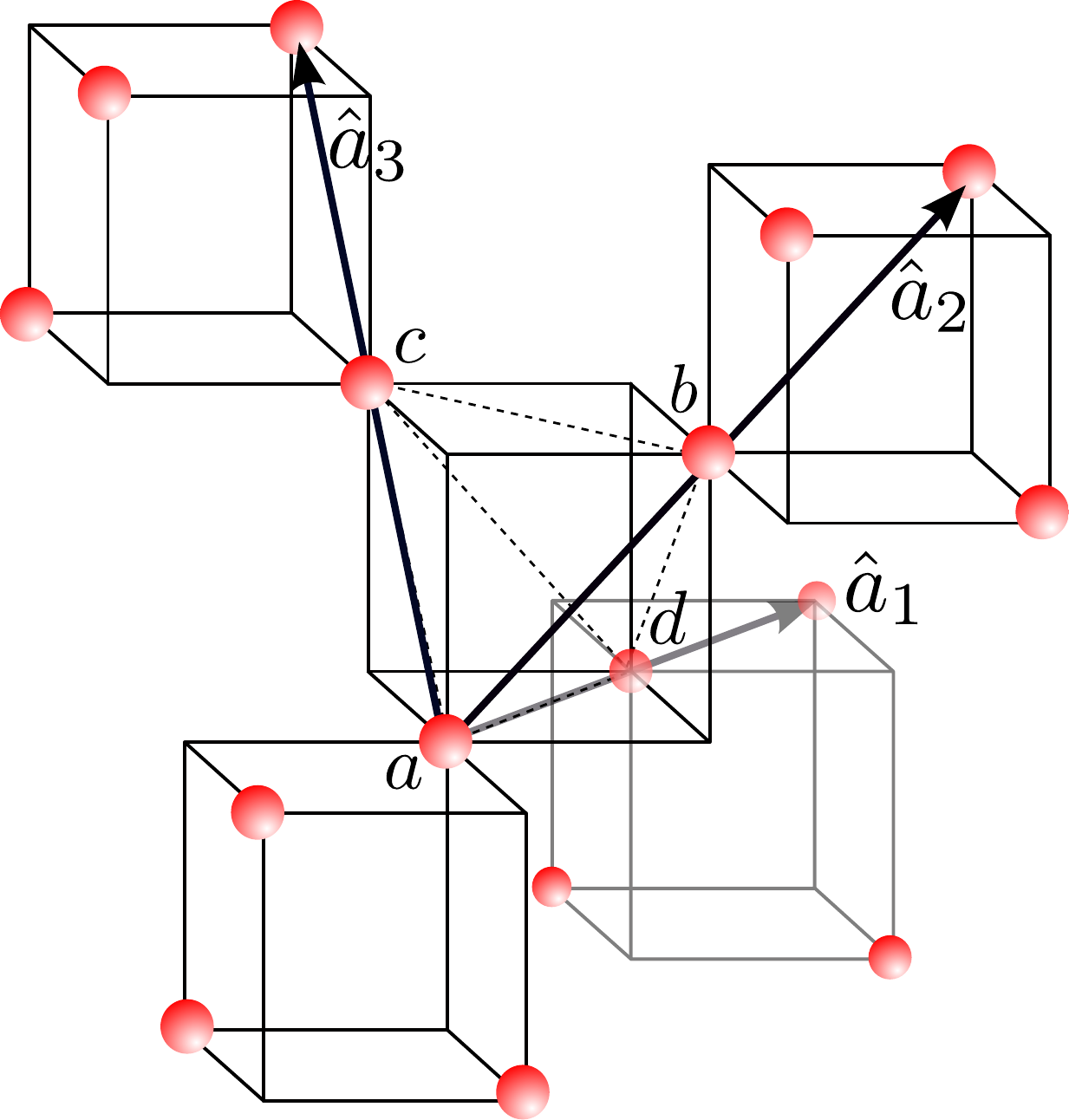}
\caption{Schematic diagram of the pyrochlore lattice with the basis vectors $\hat{a}_1, \hat{a}_2$ and $\hat{a}_3$. Red spheres represent sites with spin $S=1/2$. The four vertices of a tetrahedron are different from one another with respect to their environment, indicating four sub-lattices through the indices $a, b, c$ and $d$. The dashed lines form a tetrahedron, signalling the geometric frustration present in the system.
}
\label{fig:pyrochlore}
\end{figure}
The Hamiltonian for a system of spins on the three dimensional pyrochlore lattice with nearest neighbour (n.n.) antifferomagnetic Heisenberg exchange, a biquadratic exchange term arising from spin-lattice coupling (SLC) and an external magnetic field may be written as
\begin{eqnarray}
H= \sum_{<\vec{r}\vec{r}'>}[J \vec{S}_{\vec{r}}\cdot\vec{S}_{\vec{r}'}-J_{sl}\big(\vec{S}_{\vec{r}}\cdot\vec{S}_{\vec{r}'}\big)^2] -h\sum_{\vec{r}} S_{\vec{r}}^z~,
\label{eqn:Hamiltonian}
\end{eqnarray}
where $\vec{r}\in (\vec{R},j)$  with the lattice vector $\vec{R}=n_1\hat{a}_1+n_2\hat{a}_2+n_3\hat{a}_3$. Here, $\hat{a}_1, \hat{a}_2$ and  $\hat{a}_3$ are the three non-orthogonal basis vectors of the pyrochlore lattice, and $n_1, n_2, n_3$ are coordinate numbers along their respective basis vectors (see Fig.(\ref{fig:pyrochlore})). the four sub-lattice indices in a tetrahedron are given by the index $j\in \{a,b,c,d\}$. The coupling constant $J$ is the n.n. spin exchange coupling, while $J_{sl}>0$ measures the strength of the SLC~\cite{PhysRevLett.85.4960,PhysRevLett.88.067203,PhysRevB.66.064403, PhysRevLett.93.197203,PhysRevLett.116.257201}. The external magnetic field ($h$) is applied along the $z$-direction, and perpendicular to the plane containing the basis vectors $\hat{a}_1$ and $\hat{a}_2$. For $N_1$, $N_2$ and $N_3$ being the number of units of each sub-lattice along the $\hat{a}_1$, $\hat{a}_2$  and $\hat{a}_3$ directions respectively, the total number of sites in the lattice is $\mathcal{N}=4N_1 N_2N_3$. Below, we will consider, say, periodic boundary conditions (PBC) along $\hat{a}_1$ direction. Further, for $\delta$ denoting the distance between n.n sites, $L_{\hat{a}_1}=2\delta N_1$, $L_{\hat{a}_2}=2\delta N_2$ and $L_{\hat{a}_3}=2\delta N_3$ are the lengths along the $\hat{a}_1$, $\hat{a}_2$ and $\hat{a}_3$ directions respectively. Hereafter, we will consider $\delta=1$.
\par
For a $S=1/2$ spin system, the SLC can simplified to the following quadratic form (for details, see Appendix-\ref{AppendixA})
\begin{eqnarray}
H_{SL}=\frac{J_{sl}}{2} \sum_{<\vec{r}\vec{r}'>}~\vec{S}_{\vec{r}}\cdot\vec{S}_{\vec{r}'}~,
\label{eqn:SL1}
\end{eqnarray} 
displaying that $J_{sl}$ simply renormalises the n.n. Heisenberg exchange coupling $J$. Thus, $J_{sl}$ in the $S=1/2$ pyrochlore system further strengthens an antiferromagnetic $J>0$, whereas it competes with a ferromagnetic $J<0$. Thus, eqn.(\ref{eqn:Hamiltonian}) simplifies to
\begin{eqnarray}
H= J'\sum_{<\vec{r}\vec{r}'>} \vec{S}_{\vec{r}}\cdot\vec{S}_{\vec{r}'} -h\sum_{\vec{r}} S_{\vec{r}}^z~,
\label{eqn:Hamiltonian1}
\end{eqnarray}
where $J'=J+\frac{J_{sl}}{2}$ is the effective spin coupling constant. 
In the following section, we will derive the twist operator for the pyrochlore lattice, and employ it in formulating an OYA-like criterion for predicting magnetisation plateaux possible in this system.
\section{Twist Operators and the OYA-criterion}\label{sectionIII}
A pyrochlore lattice can be considered as a collection of parallel layers of two dimensional kagome lattices, with interpolating layers of two dimensional triangular lattices~\cite{ravi2018pyrochlore}. In Fig.(\ref{fig:pyrochlore}), we choose the kagome layers to lie in planes containing the basis vectors $(\hat{a}_{1},\hat{a}_{2})$, with the non-orthogonal basis vector $\hat{a}_{3}$ running between the parallel kagome layers. With this, we construct the twist operator for the pyrochlore lattice along, say, the direction $\hat{a}_{1}$ by using that developed recently for the kagome lattice~\cite{pal2018non}. For this, we first write down the twist operators for the four individual sub-lattices $j\in (a,b,c,d)$
\begin{align}
\hat{O}_a &= \exp \big[i\frac{2\pi}{N_1}\sum_{\vec{R}}(n_1+\frac{n_2}{2}+\frac{n_3}{2}) \hat{S}^z_{\vec{R},a}\big]~,\nonumber \\
\hat{O}_b &= \exp \big[i\frac{2\pi}{N_1}\sum_{\vec{R}}(n_1+\frac{n_2}{2}+\frac{n_3}{2}+\frac{1}{4}) \hat{S}^z_{\vec{R},b}\big]~,
\end{align}
while $\hat{O}_c \equiv \hat{O}_b$, and $\hat{O}_d$ is identical in form to $\hat{O}_b$ but with the factor of $1/4$ in the exponent being replaced by $1/2$.
As spin components at different sites commute, we can combine all four twist operators into one: $\hat{O}=\hat{O}_a\hat{O}_b\hat{O}_c\hat{O}_d$. The final form of the twist operator for the pyrochlore lattice is then
\begin{align}
\hat{O}=&\exp \big[i\frac{2\pi}{N_1}\Big(\sum_{\vec{r}}(n_1+\frac{n_2}{2}+\frac{n_3}{2}) \hat{S}^z_{\vec{r}}\nonumber\\
&+\sum_{\vec{R}}(\frac{1}{4} \hat{S}^z_{\vec{R},b}+\frac{1}{4} \hat{S}^z_{\vec{R},c}+\frac{1}{2} \hat{S}^z_{\vec{R},d})\Big)\big]~,
\label{twistoperator}
\end{align}
where $N_1$ is the number of units of a given sub-lattice along the $\hat{a}_1$ direction. We note that the terms $\exp \big[i\frac{2\pi}{N_1}\sum_{\vec{r}}\frac{n_3}{2}\Big]$  and $\exp \big[i\frac{2\pi}{N_1}\sum_{\vec{R}}(\frac{1}{2} \hat{S}^z_{\vec{R},d})\big]$ in eqn.(\ref{twistoperator}) are extra with respect to the twist operator for the kagome lattice formulated in Ref.[\cite{pal2018non}]. The first of these phases arises due to the contribution from the third non-orthogonal basis vector $\hat{a}_3$. The second phase, on the other hand, is simply due to fourth sub-lattice ($d$) of the pyrochlore system. 
\par 
Defining $\hat{T}_{\hat{a}_1}$ as a translation operator along $\hat{a}_1$ direction, such that $\hat{T}_{\hat{a}_1} \vec{S}_{n_1, n_2, n_3} \hat{T}^\dagger_{\hat{a}_1}=\vec{S}_{n_1+1, n_2, n_3}$, yields the following identity for the pyrochlore lattice (the detailed steps of which are shown in Appendix-\ref{AppendixB})
\begin{align}
\hat{T}_{\hat{a}_1}\hat{O}\hat{T}_{\hat{a}_1}^\dagger 
&= \hat{O}~\exp [-i2\pi~(4N_2 N_3)(\hat{m}-\frac{\hat{S}^z_{\boxtimes}}{4})]~,
\label{OYAcriterion}
\end{align}
where $\hat{m}=\hat{S}^z_{Tot}/(4N_1N_2N_3)$ is the magnetisation per site operator, with $\hat{S}^z_{\text{Tot}}=\sum_{\vec{r}}\hat{S}^z_{\vec{r}}$ being the total magnetisation operator, and $\hat{S}^z_{\boxtimes}$ is the $z-$component of the four spins in a tetrahedron. 
For a finite magnetic field, one can now predict the possibility of magnetisation plateaux by deriving an OYA-like criterion~\cite{PhysRevLett.78.1984} from this relation in term of the fractional magnetisation per site $m/m_s$, where $m_s=1/2$ is the saturation magnetisation per site:
\begin{eqnarray}
\frac{Q_m}{2}(\frac{m}{m_s}-\frac{S^z_{\boxtimes}}{2})=n~,
\label{OYA2}
\end{eqnarray}
where $Q_m~(=~4,16, ..$ etc.) is the magnetic unit cell and $n$ is an integer. 
Thus, for $S^z_{\boxtimes}=2$ and $Q_m=4$ (the fundamental lattice unit cell), the criterion predicts possible magnetisation plateaux at $m/m_{s}=0$ and $1/2$ for $n=-2$ and $-1$ respectively. On the other hand, 
for $Q_m=16$, plateaux at $1/8, 1/4, 3/8, 1/2, 5/8, 3/4,$ and $7/8$ are predicted. In keeping with Ref.(\cite{pal2018non}), these plateaux are likely good candidate ground states in the search for topological order in a three dimensional geometrically frustrated spin-$1/2$ system. We will see below that identical predictions are obtained for magnetisation plateaux based on arguments employing a spin-parity operator for the pyrochlore lattice.
\par
As there are some experimental indications of plateaux obtained in $S=3/2$ pyrochlore systems~\cite{PhysRevLett.94.047202,Kojima2010}, we comment briefly here on magnetisation plateaux that are similarly obtained from the twist-operator approach for this system as well. From eqn.\eqref{OYAcriterion}, with $S^z_{\boxtimes}=4$, we obtain an OYA-like criterion for $S=3/2$ system
\begin{eqnarray}
\frac{3Q_m}{2}(\frac{m}{m_s}-1)=n~,
\label{OYA3}
\end{eqnarray}
where $Q_m~(=~4,16, ..$ etc.) is the magnetic unit cell and $n$ is an integer. For $Q_{m}=4$, criterion for plateaux at $m/m_{s}=1/2, 2/3$ and $5/6$ are satisfied for $n=-3,-2$ and $-1$ respectively. Further, a plateau at $m/m_{s}=3/4$ is obtained for an extended unit cell of $Q_{m}=16$ and $n=-6$. While a plateau at $m/m_{s}=1/2$ has been verified in the spinel materials CdCr$_2$O$_4$~\cite{PhysRevLett.94.047202} and ZnCr$_2$O$_4$~\cite{Kojima2010}, there are only preliminary indications of plateaux at $m/m_{s}=2/3, 3/4$ and $5/6$ thus far~\cite{Kojima2010}.
\section{Spin-parity and magnetisation plateaux}
The spin-parity operation $S^x_{\vec{r}}\rightarrow -S^x_{\vec{r}}$,  $S^y_{\vec{r}}\rightarrow -S^y_{\vec{r}}$ and $S^z_{\vec{r}}\rightarrow S^z_{\vec{r}}$ leave the Hamiltonian (\ref{eqn:Hamiltonian1}) invariant.
The operation corresponds to a $\pi$-rotation of all spins ($\vec{S}_{i}=\frac{1}{2}\vec{\sigma}_{i}$, where $\sigma$'s are Pauli spin matrices) about the $z$-axis~\cite{PhysRevB.57.8494,jalal2016topological} and can be written as
\begin{align}
\mathcal{S} = \exp[i\frac{\pi}{2}\sum_{\vec{r}}~\sigma_{\vec{r}}^z]= \prod_{\vec{r}} ~i\sigma_{\vec{r}}^z= \mathcal{W}\times \mathcal{Z}~,
\end{align}
where $\mathcal{W}=\exp[i\frac{\pi}{2}\mathcal{N}]$ and $\mathcal{Z}=\prod_{\vec{r}} ~\sigma_{\vec{r}}^z$, with $\mathcal{N}$ the total number of sites in the lattice. Then, we can rewrite $\mathcal{Z}$ as
\begin{align}
\mathcal{Z} &= \exp[i\pi(\hat{S}_{Tot}^z-\mathcal{N} S)]~= \exp[i\pi \mathcal{N}(\hat{m}- S)]~,
\label{eqn:operatorZ}
\end{align}
where $\hat{S}_{Tot}^z=\frac{1}{2}\sum_{\vec{r}}~\sigma_{\vec{r}}^z$ is the total magnetisation operator of the system, and $\hat{m}=\frac{1}{\mathcal{N}}\sum_{\vec{r}} \hat{S}^z_{\vec{r}}$ is magnetisation per site operator. The operator $\mathcal{Z}$ is clearly a global operator, and takes values $\pm 1$ corresponds to two topologically different parity sectors of the many-body Hilbert space. It is straightforward to show that the spin-parity operator $\mathcal{Z}$ commutes with the Hamiltonian $H$: $[\mathcal{Z}, H]=0$ (see Appendix-\ref{AppendixC} for details). Thus, the eigenvalues of $\mathcal{Z}$ are good quantum numbers. Therefore, from the quantization condition of eqn.(\ref{eqn:operatorZ}), we have
\begin{eqnarray}
\mathcal{N}(m-S)=n
\label{eqn:OYA}
\end{eqnarray}
where $n$ is any integer. 
\par
Thus far, we have not invoked any notion of a specific lattice geometry in reaching eqn.\eqref{eqn:OYA}. In order to make conclusions specific to the pyrochlore lattice, we note that since the four lattice sites of a tetrahedron form the minimum unit cell of a pyrochlore lattice with periodic boundary conditions in all directions, we must impose the condition: $\mathcal{N}=4N_1N_2N_3$. For $S=1/2$, the magnetisation ($m$) values satisfying the condition eqn.(\ref{eqn:OYA}) correspond to states with a well-defined parity. If protected by a spectral gap, we expect that such states correspond to non-trivial topologically ordered spin liquid ground states and exhibit plateaux in the magnetisation vs. field plot. We will now show that, upon imposing the condition 
\begin{equation}
\mathcal{N}=4N_1N_2N_3 = q Q_m~,
\label{Ncondition}
\end{equation}
where $Q_m=4(3p+1)$ is the magnetic unit cell, $p$ belongs to the set of non-negative integers, $q$ to the set of non-zero positive integers together with $S=1/2$, eqn.(\ref{eqn:OYA}) gives the same predictions for the positions of magnetisation plateaux for pyrochlore lattice as obtained from the OYA-criterion (eqn.\eqref{OYA2}). The fundamental unit cell of the pyrochlore lattice is $Q_m=4~ (\text{for} ~p=0)$ and the simplest enlarge unit cell is $Q_m=16$ (for $p=1$). Then, we can rewrite eqn.\eqref{eqn:OYA} as
\begin{equation}
\frac{qQ_m}{2}(\frac{m}{m_s}-1) = n~,
\label{eqn:OYA11}
\end{equation}
where $m_{s}=1/2$ is the saturation magnetisation per site.
\par
We find two cases for the possible plateaux states of the minimum unit cell $Q_m=4$. First, we define $\mathcal{Z}_{Py}$ as the spin-parity operator relevant to the pyrochlore lattice, and is obtained from eqn.\eqref{eqn:operatorZ} by imposing the condition eqn.\eqref{Ncondition}. Then, for $\mathcal{Z}_{Py}=-1$, such that $n$ is an odd integer from eqn.\eqref{eqn:operatorZ}. When put into eqn.\eqref{eqn:OYA11}, this implies that $q$ is an odd integer, and thus
\begin{equation}
2(\frac{m}{m_s}-1)=2k+1~,~~~k\in~\mathrm{integer}~.
\end{equation}
For $k=-1$, this gives a possible magnetisation plateaux at $m/m_{s}=1/2$. Similarly, for $\mathcal{Z}_{Py}=1$, such that $n$ is an even integer, eqn.\eqref{eqn:OYA11} implies that if $q$ is odd 
\begin{equation}
2(\frac{m}{m_s}-1)= 2w~,~~~w\in~\mathrm{integer}~.
\end{equation}
For $w=-1$, we obtain a possible plateau at $m/m_{s}=0$. Finally, if $q$ is even integer 
\begin{equation}
2(\frac{m}{m_s}-1)= l~, ~~~l\in~\mathrm{integer}~.
\end{equation}
If $l$ is either odd or even, we obtain the $m/m_{s}=1/2$ and $m/m_{s}=0$ plateau respectively (as before). The analysis can also be extended to the case of the simplest enlarged unit cell, i.e., $Q_m=16$, obtaining possible plateaux states at $m/m_{s}= 0,~ 1/8,~ 1/4,~ 3/8,~ 1/2, ~5/8$ and $7/8$. We end by observing that the identical predictions of plateaux for the pyrochlore lattice from the analyses of the twist ($\hat{O}$) and the spin-parity operators ($\mathcal{Z}_{Py}$) arises from the following relation between the matrix elements obtained from eqn.\eqref{OYAcriterion} and eqn. \eqref{eqn:operatorZ} acting on the ground state $|\psi_{0}\rangle$ 
\begin{equation}
\langle \psi_{0}|(\hat{O}^{\dagger}\hat{T}_{\hat{a}_1}\hat{O}\hat{T}_{\hat{a}_1}^\dagger)^{\frac{N_{1}}{2}}|\psi_{0}\rangle 
= \langle\psi_{0} |\mathcal{Z}_{Py}|\psi_{0}\rangle~. 
\label{LSMparity}
\end{equation}
\section{Discussion}
In conclusion, we have predicted possible magnetisation plateaux ground states for the $S=1/2$ pyrochlore lattice with arguments based on an OYA-like criterion and the spin-parity operator. Our analysis shows that for the fundamental lattice unit cell (i.e., a tetrahedron), $m/m_{s}=0$ and $1/2$ are the two possible plateaux states, while other plateaux with fractional magnetisation arise with the enlargement of the unit cell. Similar results have been obtained for magnetisation plateaux in the spin-$1/2$ kagome system~\cite{nishimoto2013controlling,PhysRevB.88.144416,pal2018non}. We have also obtained results from the twist-operator approach for $S=3/2$ pyrochlore systems that predict plateaux at $m/m_{s}=1/2,2/3,3/4$ and $5/6$. While a plateaux at $m/m_{s}=1/2$ has been observed in certain spinels~\cite{PhysRevLett.94.047202,Kojima2010}, there are only preliminary results on plateaux at other fractions~\cite{Kojima2010}.
\par
It is important to note that our analysis does not depend on the perturbative expansion of any coupling, instead relying only on the symmetries of the Hamiltonian. Given that the operators employed in reaching these predictions are non-local (i.e., global) in nature, we expect that the properties of the corresponding ground states will be topologically distinct. For instance, some recent theoretical studies on the spin-$1/2$ kagome lattice have also revealed the topological nature of magnetisation plateaux ground states~\cite{PhysRevB.90.174409,10.1088/1367-2630/ab05ff}. It is also interesting to note that while a SLC stabilises the magnetisation plateaux of the antiferromagnetic ($J>0$) quantum pyrochlore, it will compete for the case of a ferromagnetic exchange ($J<0$). Thus, while a symmetry broken ferromagnetic ground state is to be expected for a dominant $J<0$, tuning the SLC such that the signature of the effective coupling $J'=J+\frac{J_{sl}}{2}$ changes from negative to positive can induce frustration into the system. We can then expect the appearance of magnetisation plateaux in such a case upon tuning the SLC. 
\par
To our knowledge, this is the first analytical work for the $S=1/2$ pyrochlore system that predicts the existence of plateaux in the magnetisation. It will be interesting to test these predictions numerically by looking for signatures in, for instance, exact diagonalization (ED) studies of small clusters. Extending our work to the case of spinel systems (in which both A and B sites are magnetic) should be interesting, as magnetisation plateaux~\cite{Tsurkane1601982} and spin liquid ground states~\cite{PhysRevB.84.174424,PhysRevB.96.075139} in such systems are under investigation.
Finally, it will be challenging to adapt either the functional RG method ~\cite{PhysRevX.9.011005} or the renormalisation group method used recently in studying the $m/m_{s}=1/3$ plateau of the kagome system~\cite{10.1088/1367-2630/ab05ff} to the plateaux we have predicted here for the pyrochlore.

\section*{Acknowledgement}
The authors thank A. Mukherjee, S. Patra, S. Ray, R. K. Singh, G. Dev Mukherjee, M. Oshikawa, T. Momoi, S. Pujari, R. Ganesh, and especially V. Ravi Chandra for several enlightening discussions. S. Pal acknowledge CSIR, Govt. of India and IISER Kolkata for financial support. S. L. thanks the DST, Govt. of India for funding through a Ramanujan Fellowship during which this project was initiated.
\section*{Appendix}
\appendix
\section{Spin-lattice coupling for spin-half pyrochlore}\label{AppendixA}
For $S=1/2$, one can write $\vec{S}=\frac{1}{2}\vec{\sigma}$ (where $\sigma$'s are Pauli matrices). Therefore, one can simplify the bi-quadratic term in eqn.(\ref{eqn:Hamiltonian}) as follows
\begin{align}
H_{SL}&= -J_{sl}\sum_{<\vec{r}\vec{r}'>}~\big(\vec{S}_{\vec{r}}\cdot\vec{S}_{\vec{r}'}\big)^2 \nonumber \\
&= -\frac{J_{sl}}{16}\sum_{<\vec{r}\vec{r}'>}[(\sigma^x_{\vec{r}}\sigma^x_{\vec{r}'})^2+(\sigma^y_{\vec{r}}\sigma^y_{\vec{r}'})^2+(\sigma^z_{\vec{r}}\sigma^z_{\vec{r}'})^2
\nonumber \\
&~~~+2\sigma^x_{\vec{r}}\sigma^x_{\vec{r}'}\sigma^y_{\vec{r}}\sigma^y_{\vec{r}'} +2\sigma^x_{\vec{r}}\sigma^x_{\vec{r}'}\sigma^z_{\vec{r}}\sigma^z_{\vec{r}'} 
+2 \sigma^y_{\vec{r}}\sigma^y_{\vec{r}'}\sigma^z_{\vec{r}}\sigma^z_{\vec{r}'}]\nonumber \\
&= -\frac{J_{sl}}{16}\sum_{<\vec{r}\vec{r}'>}[3-2(\sigma^z_{\vec{r}}\sigma^z_{\vec{r}'} +\sigma^y_{\vec{r}}\sigma^y_{\vec{r}'} +\sigma^x_{\vec{r}}\sigma^x_{\vec{r}'})]\nonumber \\
&= - \sum_{<\vec{r}\vec{r}'>}[\frac{3J_{sl}}{16}-\frac{J_{sl}}{2}~ \vec{S}_{\vec{r}}\cdot\vec{S}_{\vec{r}'}]~.
\end{align}
Here, we used the fact that Pauli matrices at different sites commute with one another, and $\sigma^\alpha \sigma^\beta=i\epsilon^{\alpha\beta\gamma}\sigma^\gamma$, with $\{\alpha,\beta,\gamma\} \in \{x,y,z\}$ in cyclic permutation and $\epsilon^{\alpha\beta\gamma}$ is an antisymmetric tensor. In eqn.(\ref{eqn:SL1}), we have neglected a constant term.
\section{LSM calculation}\label{AppendixB}
Here, we present a calculation of the non-commutativity between twist and translation operators defined in section \ref{sectionIII}
\begin{widetext}
\begin{align}
\hat{T}_{\hat{a}_1}\hat{O}\hat{T}_{\hat{a}_1}^\dagger &= \hat{T}_{\hat{a}_1}~\exp \big[i\frac{2\pi}{N_1}\Big(\sum_{\vec{r}}(n_1+\frac{n_2}{2}+\frac{n_3}{2}) \hat{S}^z_{\vec{r}}+\sum_{\vec{R}}(\frac{1}{4} \hat{S}^z_{\vec{R},b}+\frac{1}{4} \hat{S}^z_{\vec{R},c}+\frac{1}{2} \hat{S}^z_{\vec{R},d})\Big)\big]~\hat{T}_{\hat{a}_1}^\dagger \nonumber\\
&=\hat{T}_{\hat{a}_1}~\exp[i\frac{2\pi}{N_1} \sum_{n_2,n_3,j}\Big(\hat{S}_{(1,n_2,n_3),j}^z
+2\hat{S}_{(2,n_2,n_3),j}^z+...+(N_1-1)\hat{S}_{(N_1-1,n_2,n_3),j}^z + N_1 \hat{S}_{(N_1,n_2,n_3),j}^z\Big)]~\hat{T}_{\hat{a}_1}^\dagger \nonumber\\
&~~~~\exp \big[i\frac{2\pi}{N_1}\Big(\sum_{\vec{r}}(\frac{n_2}{2}+\frac{n_3}{2}) \hat{S}^z_{\vec{r}}+\sum_{\vec{R}}(\frac{1}{4} \hat{S}^z_{\vec{R},b}+\frac{1}{4} \hat{S}^z_{\vec{R},c}+\frac{1}{2} \hat{S}^z_{\vec{R},d})\Big)\big]
\nonumber\\
&=\exp[i\frac{2\pi}{N_1} \sum_{n_2,n_3,j}\Big(\hat{S}_{(2,n_2,n_3),j}^z+2\hat{S}_{(3,n_2,n_3),j}^z+...+(N_1-1)\hat{S}_{(N_1,n_2,n_3),j}^z+N_1\hat{S}_{(N_1+1,n_2,n_3),j}^z\Big)]
\nonumber\\
&~~~~\exp \big[i\frac{2\pi}{N_1}\Big(\sum_{\vec{r}}(\frac{n_2}{2}+\frac{n_3}{2}) \hat{S}^z_{\vec{r}}+\sum_{\vec{R}}(\frac{1}{4} \hat{S}^z_{\vec{R},b}+\frac{1}{4} \hat{S}^z_{\vec{R},c}+\frac{1}{2} \hat{S}^z_{\vec{R},d})\Big)\big]
\nonumber\\
&= \hat{O}~\exp[i\frac{2\pi}{N_1} \sum_{n_2,n_3,j} N_1\hat{S}_{(1,n_2,n_3),j}^z] \exp[-i\frac{2\pi}{N_1} \sum_{\vec{r}}\hat{S}^z_{\vec{r}}]\nonumber\\
&=  \hat{O}~\exp[i2\pi \sum_{n_2,n_3,j} \hat{S}_{(1,n_2,n_3),j}^z]\exp[-i\frac{2\pi}{N_1} \hat{S}^z_{\text{Tot}}]\nonumber\\
&= \hat{O}~\exp [-i\frac{2\pi}{N_1}(\hat{S}^z_{\text{Tot}}-N_1N_2N_3\hat{S}^z_{\boxtimes})]~,
\end{align}
\end{widetext}
where $\hat{S}^z_{\boxtimes}=\hat{S}^z_{(1,1,1),a}+\hat{S}^z_{(1,1,1),b}+\hat{S}^z_{(1,1,1),c}+\hat{S}^z_{(1,1,1),d}$ is the vector sum of the $z$-component of the four sub-lattice spins in a boundary tetrahedron ($n_1\equiv 1$ corresponds to a boundary spin). $\hat{S}^z_{\text{Tot}}=\sum_{\vec{r}}\hat{S}^z_{\vec{r}}$ is the total magnetisation operator, and we have used periodic boundary conditions along $\hat{a}_1$ such that site $N_1+1\equiv$ site $1$.
\section{Commutation relation}\label{AppendixC}
We present here the calculation of the commutation relation $\big[\mathcal{Z}, H\big]$, which is equivalent to calculating the two commutators, $\big[\prod_{\vec{\tilde{r}}} \sigma_{\vec{\tilde{r}}}^z, ~\sigma_{\vec{r}}^z \big]$ and $\big[\prod_{\vec{\tilde{r}}} \sigma_{\vec{\tilde{r}}}^z,~ \vec{\sigma}_{\vec{r}}\cdot\vec{\sigma}_{\vec{r}'}  \big]$. Of these, the first commutation relation is exactly zero, as it involves only the $z$-component of Pauli matrices. The second commutator can be computed as follows
\begin{eqnarray}
&&\Big[\prod_{\vec{\tilde{r}}} \sigma_{\vec{\tilde{r}}}^z, ~\sigma_{\vec{r}}^x\sigma_{\vec{r}'}^x \Big]\nonumber \\
&=& \Big[\prod_{\vec{\tilde{r}}} \sigma_{\vec{\tilde{r}}}^z, \sigma_{\vec{r}}^x\Big]\sigma_{\vec{r}'}^x +\sigma_{\vec{r}}^x \Big[\prod_{\vec{\tilde{r}}} \sigma_{\vec{\tilde{r}}}^z, \sigma_{\vec{r}'}^x\Big] \nonumber \\
&=& \Big(0+...+ \sigma^z_{\vec{r}_1}...\Big[\sigma^z_{\vec{r}},
\sigma^x_{\vec{r}} \Big]...\sigma^z_{\vec{r}_{\mathcal{N}}}+...+0\Big)\sigma_{\vec{r}'}^x\nonumber \\
&&+ \sigma^x_{\vec{r}}\Big(0+...+ \sigma^z_{\vec{r}_1}...\Big[\sigma^z_{\vec{r}'},
\sigma^x_{\vec{r}'} \Big]...\sigma^z_{\vec{r}_{\mathcal{N}}}+...+0\Big)\nonumber \\
&=& \sigma^z_{\vec{r}_1}...(i\sigma^y_{\vec{r}})...(i\sigma^y_{\vec{r}'})...\sigma^z_{\vec{r}_{\mathcal{N}}} + \sigma^z_{\vec{r}_1}...(-i\sigma^y_{\vec{r}})...(i\sigma^y_{\vec{r}'})...\sigma^z_{\vec{r}_{\mathcal{N}}} \nonumber \\
&=&~0~.
\end{eqnarray}
In the above, we have used the following Pauli matrix identities:
$[ \sigma_n^\alpha, \sigma_m^\beta]= i\epsilon^{ \alpha\beta\gamma} \delta_{nm}\sigma_n^\gamma $, where $\{ \alpha, \beta, \gamma \}\in \{x, y, z \}$, $\epsilon^{ \alpha\beta\gamma}$ is the antisymmetric tensor and $\delta_{nm}$ is the delta function between site $n$ and $m$. In the  same way, we find that
\begin{eqnarray}
\Big[\prod_{\vec{\tilde{r}}} \sigma_{\vec{\tilde{r}}}^z, ~\sigma_{\vec{r}}^y\sigma_{\vec{r}'}^y \Big]=\Big[\prod_{\vec{\tilde{r}}} \sigma_{\vec{\tilde{r}}}^z, ~\sigma_{\vec{r}}^z\sigma_{\vec{r}'}^z \Big]=~0~.
\end{eqnarray}
Thus, we can conclude that
\begin{equation}
\big[\prod_{\vec{\tilde{r}}} \sigma_{\vec{\tilde{r}}}^z,~ \vec{\sigma}_{\vec{r}}\cdot\vec{\sigma}_{\vec{r}'}  \big]=~0~.
\label{eqn:app2}
\end{equation}
Bringing together both commutators, we find that
\begin{eqnarray}
\Big[\mathcal{Z},~ H\Big]=~0~,
\end{eqnarray}
i.e., the spin-parity operator commutes very generally with the Hamiltonian for the $S=1/2$ Heisenberg antiferromagnet in a field. 
\bibliography{reference_pyrochlore}
\end{document}